Correspondence and requests for materials should be addressed to D.F. (ddsfu@ipc.shizuoka.ac.jp)

# High-$T_c$ BaTiO$_3$ ferroelectric films with frozen negative pressure states

Desheng Fu[1,*], Kouhei Fukamachi[2], Naonori Sakamoto[2], Naoki Wakiya[2], Hisao Suzuki[2,3], Mitsuru Itoh[4], Takeshi Nishimatsu[5]

[1]Division of Global Research Leaders, Shizuoka University, Johoku 3-5-1, Naka-ku, Hamamatsu 432-8561, Japan

[2]Department of Materials Science and Chemical Engineering, Shizuoka University, Johoku 3-5-1, Naka-ku, Hamamatsu 432-8561, Japan

[3]Graduate School of Science and Technology, Shizuoka University, Johoku 3-5-1, Naka-ku, Hamamatsu 432-8561, Japan

[4]Materials and Structures Laboratory, Tokyo Institute of Technology, 4259 Nagatsuta, Yokohama 226-8503, Japan

[5]Institute for Materials Research (IMR), Tohoku University, Sendai 980-8577, Japan

**We report that an energetic plasma process is extremely effective in enlarging the unit cell volume and ferroelectric distortion of the ferroelectric oxides, resulting in a significant increase in its $T_c$. We demonstrate experimentally that $c$-axis oriented BaTiO$_3$ films can be deposited directly on quartz glass and Si substrates using such a process and that the material shows an approximately 5% expansion of its unit cell volume and approximately 4 times the ferroelectric tetragonal distortion of the bulk crystals. Such a frozen negative pressure results in a $T_c$ value that is approximately 580 K higher than that of bulk single crystals, providing a wide range of operating temperatures for the devices. The present results suggest an approach to producing ferroelectric oxides with unique properties that might be extended to ferromagnetic or superconductor oxides and demonstrate a route to a lead-free ferroelectric oxide for capacitive, ferroelectric memory, and electro-optical devices.**

Ferroelectric is crystal in which electric polarization due to displacements of charged ions are formed spontaneously and are able to be switched by an electric field. Such a spontaneous polarization allows excellent coupling effects with electric or magnetic field, mechanical force, thermal stimulus, and photons, finding applications in electronic, electromechanical and electro-optical devices [1,2]. Ferroelectricity is determined by a delicate balance between the long-range

Coulomb force and short-range repulsions and is thus extremely sensitive to chemistry, defects, electrical boundary conditions and pressure[3]. Adjusting the ferroelectric transition temperature ($T_c$) is traditionally accomplished in solid solutions by chemical substitution. However, such tuning is not without limitations. The $T_c$ of the solid solution is generally limited within the ranges of values of their parent compounds. An alternative means of adjusting $T_c$ is to use a substrate that induces a strain in the thin film. In practice, it has been shown that epitaxial strain can be used to increase $T_c$ by hundreds of degrees using molecular beam epitaxy and pulsed-laser deposition on specially selected substrates [4-6]. However, the synthesis of uniformly strained ferroelectric films is challenging. To impart strain on ferroelectric films, special single-crystal substrates that are structurally, chemically, and thermally compatible with the deposited materials and have appropriate lattice constants to impart reasonable strain on coherent films are desired. In addition, the thickness of the films must be controlled below a critical value; otherwise, undesirable relaxation towards a zero-strain state wills occur[5].

The following question arises: is there an alternative approach to enhance the $T_c$ of the ferroelectric materials? Our experimental discovery provides a promising answer to this question. In our investigations on $BaTiO_3$ and related perovskite oxides, we have discovered an unexpected volume expansion of the materials deposited by an energetic plasma process in RF magnetron sputtering. These films were observed to have a higher $T_c$ than that of bulk single crystals. The volume expansion is believed to be due to the non-equilibrium state produced by the energetic plasma process[7] and a *frozen* negative-pressure state in which ferroelectricity is enhanced[8].

Magnetron sputtering is an energetic plasma process, in which the film growth surface is under constant bombardment by an energetic particle flux as it receives the desired material for film growth[7]. In general, the particle energy is in the range of 10-100 eV, significantly higher than the traditional processing temperature of 0.1 eV ($\approx$1200 K).

While under bombardment by the energetic particles, the surface layers of the crystal are in a nonequilibrium state. The modified layers retain a memory of the conditions of their origin and exhibit unusual properties. It is not surprising that films formed using such specific processes exhibit unexpected behaviours.

In our work, BaTiO$_3$ films were prepared using RF magnetron sputtering[9] on substrates of Si and quartz glass. The former is the workhorse of semiconductor technology and has been widely explored for the integration of ferroelectric thin films[5,10], whereas the latter is transparent and is thus appropriate for optical applications. To apply an electric field to ferroelectric films, we also deposited films on (001) Si substrates coated with conducting LaNiO$_3$ films prepared by chemical solution deposition[11]. Figure 1 shows the X-ray diffraction patterns for the BaTiO$_3$ films that were deposited under the same conditions as described in the methods for these substrates. It is clear that we have successfully obtained oriented films using RF magnetron sputtering. Strikingly, the glass substrates also allowed orientational control of the film growth. Principally, an amorphous substrate cannot offer a periodic lattice for the oriented growth of crystals through a lattice-matching mechanism. The driving force for such orientational control remains unclear at present. A likely cause is the electric field applied during the RF magnetron sputtering. In particular, the ferroelectric materials have spontaneous electric polarisation due to dipoles formed by the displacement of charged ions inside the crystal unit cell. Such dipoles, if they exist, are subjected to the Coulomb forces of the *ac* electric field applied along the normal direction of the substrate, which favours the *c*-axis (polarisation axis) growth of ferroelectric films.

Another striking feature of the obtained BaTiO$_3$ films is that they exhibit great elongation along the *c*-axis of the lattice. As shown in Fig. 1a, such an elongation results in a large shift of the *00l*-reflections in comparison with the bulk single crystals. The *c*-axis lattice constants that are calculated from the *002*-reflection are summarised in Fig. 1 together with the *a*-axis lattice constants that are evaluated from the in-plane diffraction measurements. In association with the large elongation of the *c*-axis lattice, a slight increase in the *a*-axis lattice (Fig. 1b & 1c) is also observed in the BaTiO$_3$ films. Both effects directly result in an expansion of the unit cell. Under the condition described in the method, the unit cell of the film has a volume expansion of ~5% over that of bulk crystals[12]. At the same time, we observed a great enhancement in the tetragonality (*c/a*-1) of the BaTiO$_3$ films. The tetragonality of the deposited films is approximately four times that of bulk single crystals. Our results thus provide strong support for the theoretical prediction that a large tetragonal strain can be induced in ferroelectric PbTiO$_3$ by the application of a negative hydrostatic pressure, and the same effects are expected for BaTiO$_3$ in the first-principle investigations of Tinte et al.[8]. The application of negative pressure is generally believed to be experimentally infeasible, but our findings clearly indicate that a *negative* pressure state can be realised using an energetic plasma process such as RF magnetron sputtering. We note that the RF power can be used effectively to tune the lattices. When we reduced the power, we observed a reduction of the *c*-axis elongation. A similar effect was also realised by controlling the chamber pressure. Other potential approaches might involve adjusting the distance between the target and substrate because this distance, in addition to the power and pressure, can be used to tune the energy of the bombardment particle and thus tune the states of the films.

The great increase in lattice strain will lead to a more stable ferroelectric structure and result in an enhancement of the transition temperature of the paraelectric-ferroelectric phase transition. This has been demonstrated in chemical substitution[13-15] and strained expitaxial thin films [4-6]. To sense such a structural transition in our films with a *negative* pressure state, we performed temperature-dependent x-ray diffraction measurements of the out-of-plane lattices, a technique that has been widely used to determine the $T_c$ of ferroelectric thin films [4,5]. We traced the evolution of the x-ray diffraction patterns in a temperature range of 300-1260 K using a temperature step of 20 K in the heating process. Typical results are shown in Fig. 2. Here, we only show the results of BaTiO$_3$/Si films, in which the *004*-reflection of Si can act as a reference to reveal the structural evolution, but similar results were also obtained with other films. From Fig. 2, we can clearly see that there is a dramatic change in the *002*-reflection of the BaTiO$_3$ films around 980 K. This contrasts with the monotonous shift of the *004*-reflection of Si due to slight thermal expansion. Such a sharp change in the lattice reflections indicates that a structural transformation occurs at $T=T_c\sim$980 K in the BaTiO$_3$ films. The out-of-plane lattice constant calculated from this reflection is also given in Fig. 2 for comparison with those of bulk single crystals and the Si substrate. In BaTiO$_3$ bulk, the paraelectric-ferroelectric phase transition occurs at $T_c$=398 K (Ref. 16), above which temperature the lattice changes to a cubic structure and merely shows monotonously thermal expansion. The out-of-plane lattice constant (4.065 Å) of the film at 1200 K is very close to the value of $a$=4.048 Å of the bulk, indicating that the expanded BaTiO$_3$ films undergo a structural transition at $T_c\sim$980 K in which their structure changes to the normal cubic state of bulk crystals. It should be noted that such a structural change is irreversible. After the transformation, the films show both *200*- and *002*-reflections with positions close to those of bulk crystals when they are cooled to room temperature. This indicates that the films revert to the normal state of BaTiO$_3$.

To examine the ferroelectricity of the BaTiO$_3$ films, we also performed ferroelectric and dielectric measurements on BaTiO$_3$/LaNiO$_3$/Si films using capacitors formed between the Pt top-electrode and the LaNiO$_3$ bottom electrode (Fig. 3). The *D-E* hysteresis loop in Fig. 3d clearly shows the ferroelectricity of our BaTiO$_3$ films. The electrical insulation of these films is very excellent and can allow us to apply voltage over 200 V to the 600-nm films. These capacitors with low dielectric loss also allowed us to observe the dielectric behaviours over a wide temperature range of 2-950 K for the first time. The dielectric behaviours suggest that the tetragonal structure of the BaTiO$_3$ films is stable even at the lowest temperature used in our measurements. Within the large temperature range of 2-600 K, the BaTiO$_3$ films show nearly complete temperature independence of the dielectric constant; it maintains a value close to that of the *c*-axis single crystals (Fig. 3a & 3c)[17]. At the same time, the dielectric loss within this temperature range is low. These two features are of great significance for technological applications. With further heating of the sample, a small shoulder was observed around 700 K. This shoulder was strongly dependent on the measurement frequency and thus can be reasonably attributed to space charge polarisation, likely due to defects or electrode interface effects[18]. Upon further heating, we observed an inflection point of the dielectric constant around 900 K, at which point it began to steadily decrease with temperature. This inflection point is approximately 80 K lower than the $T_c$ observed in the X-ray measurements. A similar inflection point also occurs in the lattice constants for $T<T_c$, as shown in Fig. 2b. Although the present measurements did not allow us to observe the dielectric behaviour for temperatures $T>T_c$, our dielectric measurements unambiguously show that our BaTiO$_3$ films have a greater shift in paraelectric-ferroelectric phase transition than bulk single crystals, providing strong support for the results obtained from the X-ray diffraction measurements.

To gain further insight into the origins of this effect, we conducted first-principle density functional theory calculations within the Wu and Cohen generalised gradient approximation (GGA)[19]. We performed full, unconstrained optimisation of the structural parameters of tetragonal BaTiO$_3$ as a function of external pressure ranging from 0 to -6 GPa. At a given pressure, the set of parameters that minimised the enthalpy was determined. The Wu and Cohen GGA function allows us to estimate the structural parameters of BaTiO$_3$ (Ref. 20). The results are summarised in Fig. 4. The potential well for atomic displacement along the tetragonal direction becomes much deeper when the negative pressure increases, clearly indicating that the negative pressure stabilises the ferroelectric tetragonal phase. We found that the effect of negative isotropic pressure on the cell shape is to stretch the unit cell along the $c$ axis and to slightly squeeze it in the plane. As a result, the lattice strain and the cell volume increase with negative isotropic pressure. As shown in Fig. 4c, the -5 GPa pressure expands the theoretical cell from 64.13 Å$^3$ to 67.50 Å$^3$. Correspondingly, the well depth changes from 17.54 meV (~203.6 K) to 59.24 meV (~687.6K). We can use this energy change to predict the shift in $T_c$, estimated to be approximately 484 K. Experimentally, we found that the cells are expanded to a value of approximately 67.6 Å$^3$ from the bulk crystal volume of 64.34 Å$^3$. Correspondingly, $T_c$ is shifted to a temperature approximately 580 K higher than the bulk BaTiO$_3$ crystal $T_c$ of approximately 398 K (Ref. 16). Therefore, we may say that the first-principles calculations effectively predict the enhancement of the ferroelectric $T_c$.

In summary, we have demonstrated that a negative pressure state can be *frozen* in prototypical ferroelectric BaTiO$_3$ by an energetic plasma process, resulting in a large ferroelectric distortion and, thus, a great enhancement in the ferroelectric phase transition. Our finding opens a new avenue to tuning the physical properties of ferroelectric materials and might be extended to explore novel phenomena produced by negative pressures in various oxides, for examples, ferromagnetic[21] or superconductor oxides [22], which have thus far been inaccessible. Moreover, our *c*-axis oriented BaTiO$_3$ films in direct contact with transparent glass and Si substrates provide promising applications in hybrid ferroelectric-semiconductor devices and high-speed electro-optical devices, in which the sizeable electro-optical coefficients of BaTiO$_3$ can be put to practical use over a broad range of operating temperatures.

Methods

We used a RF magnetron sputtering system from the SEED LAB Corporation to prepare the BaTiO$_3$ films. The deposition chamber was first evacuated to below $2\times10^{-5}$ Pa and then filled with a mixture of Ar and O$_2$ gases (Ar/O$_2$=8.0:2.0, total flow rate 10 sccm) to a pressure of 2 Pa. The film was deposited at a substrate temperature of 773 K using an input RF power of 150 W that was applied between the targets and the substrates (the distance between them was approximately 4.5 cm). The target has a diameter of 5 cm, allowing for the growth of thin films with large areas. A two-hour deposition leads to the growth of ferroelectric films with thicknesses of approximately 600 nm.

The out-of-plane x-ray diffraction measurements were performed using a power x-ray diffractometer, MXP-18XHF-SRA (MAC SCIENCE), with Cu-*K*α radiation and a temperature controller for use in the high-temperature measurements. The in-plane x-ray diffractions were conducted using a Smartlab Multipurpose XRD system from

Rigaku with Cu-*K*α radiation. We used a Physical Property Measurement System from Quantum Design combined with a Hewlett Packard 4284A precision LCR meter to obtain the temperature behaviours of the dielectric constant in the temperature range of 2-350 K. For higher temperatures, a homemade system was used. Dielectric data were also collected upon heating. The *D-E* ferroelectric loops were measured using a ferroelectric measurement system from the TOYO Corporation.

The details of the first principles calculations can be found in Ref. 20. All calculations were performed with the ABINIT code[23,24]. A recently developed generalised gradient approximation (GGA) functional of Wu and Cohen was used to determine the total-energy surfaces for zone-centre distortions in the tetragonal structure and its structural parameters. Because the volume $V$ changes under pressure $p$, we use the enthalpy $H = E + pV$ rather than the total energy $E$ to compare the difference between the cubic and tetragonal structures.

Acknowledgments We acknowledge the support of a Grant-in-Aid for Scientific Research from MEXT, Japan. Computational resources were provided by the Center for Computational Materials Science, the Institute for Materials Research, Tohoku University and the Next Generation Super Computing Project, Nanoscience Program, MEXT, Japan.

Author contribution statement D.F. designed the experiments, performed the X-ray diffraction and dielectric measurements, and wrote the paper. K. F. prepared samples and measured the ferroelectric loop. N. F., N. W., and H.S. contributed to the sample preparation and ferroelectric measurements. M. I. contributed to the x-ray diffraction and dielectric measurements. T. N. performed the first-principles calculations. All authors discussed the results and commented on the manuscript.

Additional information. The authors declare no competing financial interests.

Figure captions

Figure 1. *c*-axis-oriented BaTiO$_3$ deposited using an energetic plasma process in RF magnetron sputtering. (a) Out-of-plane and (b) in-plane x-ray diffraction patterns for films grown on quartz glass, Si, and LaNiO$_3$(LNO)/Si substrates. (c) The lattice parameters of the films compared with those of bulk single crystals. The *c*-axis lattice constants were calculated from the *002*-reflection in Fig. 1a, and the *a*-axis lattice constants were evaluated from the in-plane diffractions given in Fig. 1b. In contrast to the out-of-plane *c*-axis orientation, there is no preferentially oriented in-plane growth of crystals. This can be clearly seen from the in-plane x-ray diffraction patterns, in which only (*hk*0) reflections from the *a*-domains were observed.

Figure 2. Temperature evolution of the out-of-plane lattices of the BaTiO$_3$ films. (a) The variation in the 002/200-reflections of BaTiO$_3$ films deposited directly on Si and the 400-reflection of the Si substrate. (b) The lattice constants obtained from the 002/200-reflections can be compared with those of BaTiO$_3$ crystals. A structural change is clearly seen at $T=T_c\sim 980$ K in the film.

Figure 3. Dielectric and ferroelectric behaviours of the BaTiO$_3$ films. (a) Dielectric constant and (b) dielectric loss of Pt/BaTiO$_3$/LaNiO$_3$/Si films. For comparison, the *c*-axis dielectric constants of the BaTiO$_3$ single crystal (Merz, Ref. 17) are shown in (c). The *D-E* ferroelectric hysteresis loop obtained at room temperature is shown in (d).

Figure 4. Negative pressure effects in BaTiO$_3$. (a) Total-energy surfaces for atomic displacement *u* along the [001] direction. (b) Calculated lattice constants and tetragonality *c/a* under different negative pressures. (c) Calculated volume. (d) Change in potential well depth. The right axis gives units in Kelvin and can be used to evaluate the ferroelectric transition temperature.

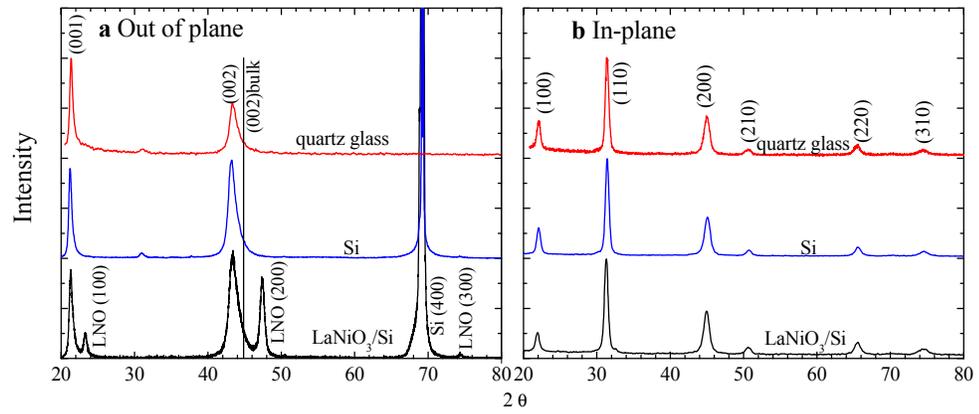

| c | Lattice parameters at room temperature | | | |
|---|---|---|---|---|
| | $a$ (Å) | $c$ (Å) | $(c-a)/a$ (%) | $V$ (Å$^3$) |
| BaTiO$_3$/glass | 4.028 | 4.169 | 3.5 | 67.64 |
| BaTiO$_3$/Si | 4.018 | 4.188 | 4.2 | 67.61 |
| BaTiO$_3$/LaNiO$_3$/Si | 4.023 | 4.177 | 3.8 | 67.60 |
| BaTiO$_3$ crystals | 3.9930 | 4.0356 | 1.1 | 64.34 |

Figure 1

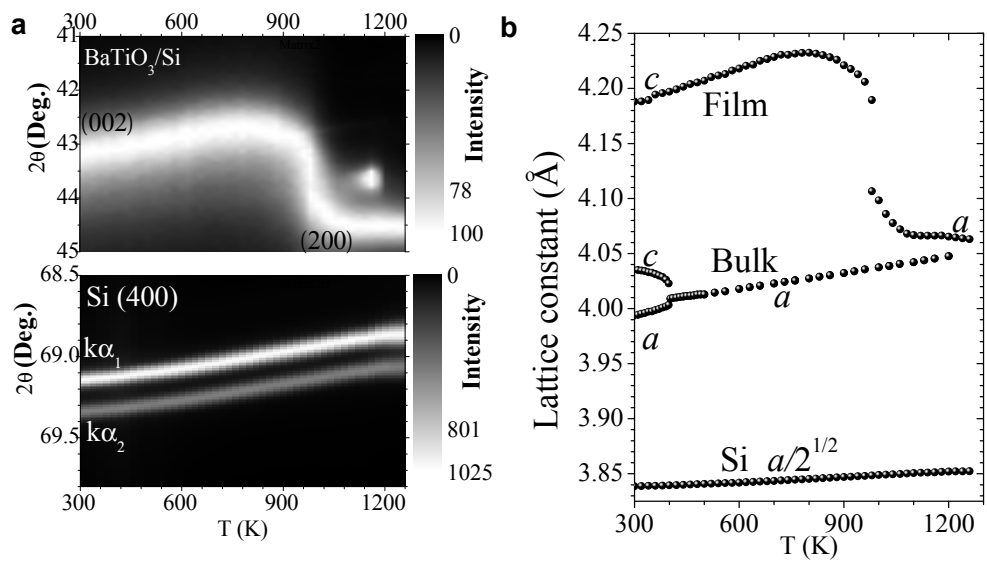

Figure 2

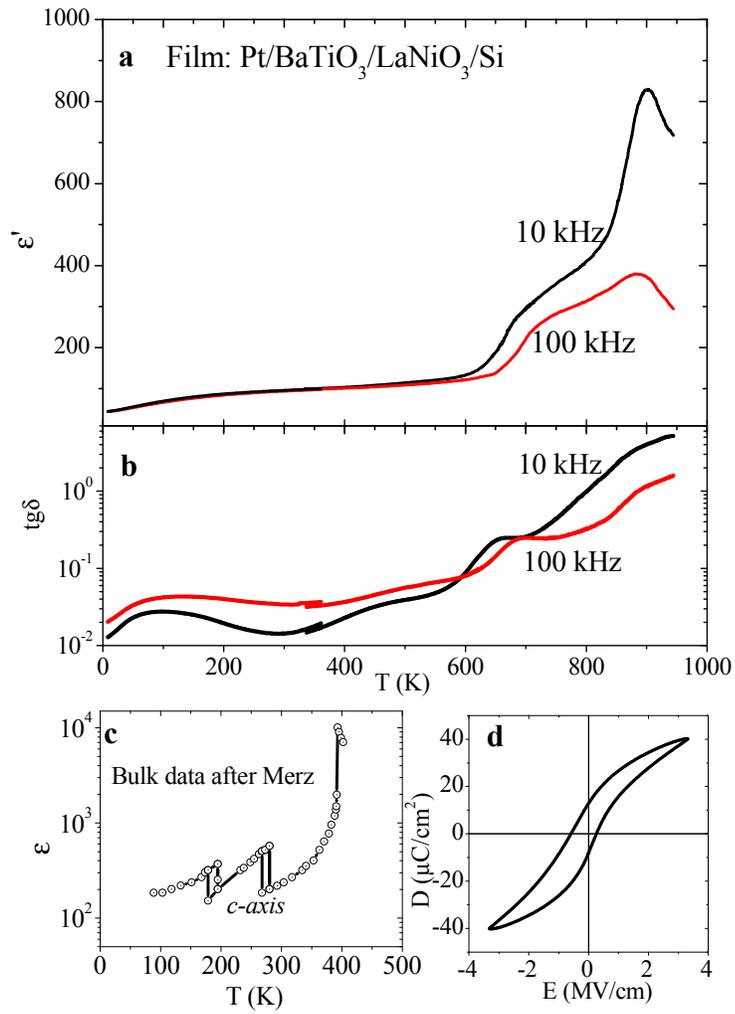

Figure 3

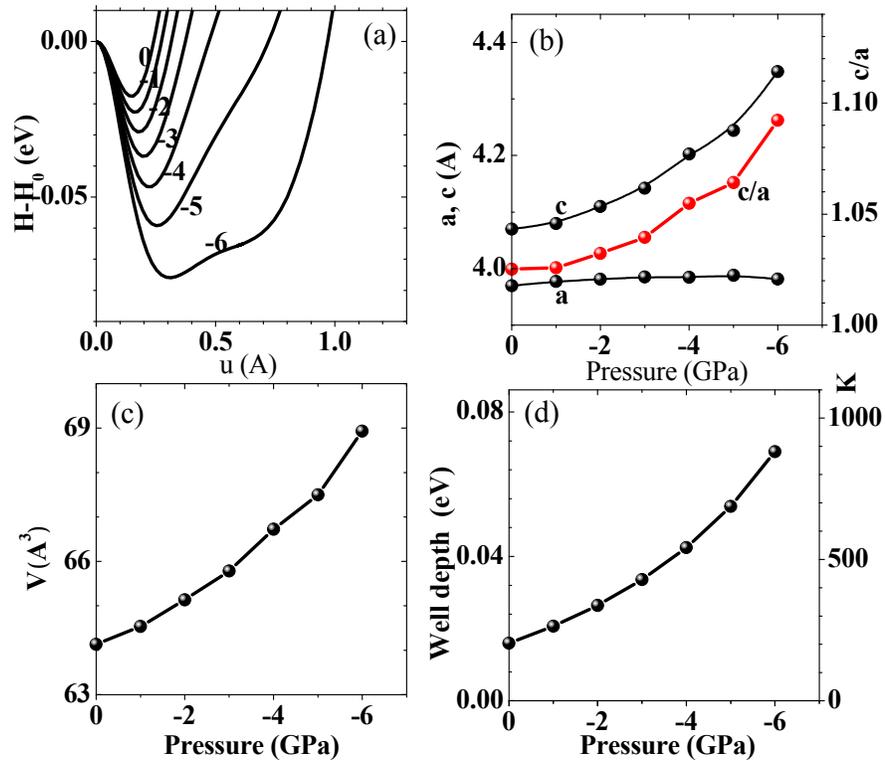

Figure 4